\documentclass[preprint]{aastex}
\usepackage{color}
\usepackage{natbib}
\usepackage{multirow}

\begin{document}
\title{ALMA and IRIS Observations of the Solar Chromosphere I: an On-Disk Type II Spicule}

\author{Georgios Chintzoglou\altaffilmark{1,2}, Bart De Pontieu\altaffilmark{1,4,5}, Juan Mart\'inez-Sykora\altaffilmark{1,3,4}, Viggo Hansteen\altaffilmark{1,3,4,5}, Jaime de la Cruz Rodr\'iguez\altaffilmark{6}, Mikolaj Szydlarski\altaffilmark{4,5}, Shahin Jafarzadeh\altaffilmark{4,5}, Sven Wedemeyer\altaffilmark{4,5}, Timothy S. Bastian\altaffilmark{7} and Alberto Sainz Dalda\altaffilmark{1,3,8}}

\email{gchintzo@lmsal.com}

\altaffiltext{1}{Lockheed Martin Solar \& Astrophysics Laboratory, Palo Alto, CA 94304, USA}
\altaffiltext{2}{University Corporation for Atmospheric Research, Boulder, CO 80307-3000, USA}
\altaffiltext{3}{Bay Area Environmental Research Institute, NASA Research Park, Moffett Field, CA 94035, USA}
\altaffiltext{4}{Rosseland Center for Solar Physics, University of Oslo, P.O. Box 1029 Blindern, NO0315, Oslo, Norway}
\altaffiltext{5}{Institute of Theoretical Astrophysics, University of Oslo, P.O. Box 1029 Blindern, NO0315, Oslo, Norway}
\altaffiltext{6}{Institute for Solar Physics, Department of Astronomy, Stockholm University, AlbaNova University Centre, SE-106 91, Stockholm, Sweden}
\altaffiltext{7}{National Radio Astronomy Observatory, 520 Edgemont Road, Charlottesville, VA 22903, USA}
\altaffiltext{8}{Stanford University, HEPL, 466 Via Ortega, Stanford, CA 94305-4085}

\begin{abstract}
We present observations of the solar chromosphere obtained simultaneously with the Atacama Large Millimeter/submillimeter Array (\emph{ALMA}) and the Interface Region Imaging Spectrograph (\emph{IRIS}). The observatories targeted a chromospheric plage region of which the spatial distribution (split between strongly and weakly magnetized regions) allowed the study of linear-like structures in isolation, free of contamination from background emission. Using these observations in conjunction with a radiative magnetohydrodynamic 2.5D model covering the upper convection zone all the way to the corona that considers non-equilibrium ionization effects, we report the detection of an on-disk chromospheric spicule with \emph{ALMA} and confirm its multithermal nature. 

\end{abstract}


\section{Introduction}\label{intro}

Chromospheric spicules were discovered in the 1870s in wide-slot spectroscopic observations in H$\alpha$ by A. Secchi (he called them in French \emph{petits filets}, i.e., ``little strings'', or \emph{poils}, i.e., ``fur'', due to their fine and slender appearance; \citealt{Secchi_1877}). We now know that spicules are jets of chromospheric material that are seen as rooted at the chromospheric network. A new class of spicules, termed as ``Type-II spicules'', was found a little over a decade ago in high-resolution imaging observations taken at the \ion{Ca}{2} H line \citep{DePontieu_etal_2007b}. These are more slender (apparent widths $\lesssim1\arcsec$) and exhibit higher plane-of-the-sky speeds ($\approx$50$-$100\,km s$^{-1}$) than their ``traditional'' counterparts. Their lifetimes differ, depending on whether we observe them in low chromospheric temperatures (e.g., in \ion{Ca}{2} H; $\Delta t\approx$ 10-150\,s) or high chromospheric or transition region temperatures (e.g., \ion{Mg}{2} h\&k and \ion{Si}{4}; $\Delta t\approx$ 3-10\,mins; \citealt{Pereira_etal_2014, Skogsrud_etal_2016}). Type-II spicules have been proposed as contributing to the heating of the corona based on observational studies \citep{DePontieu_etal_2009, DePontieu_etal_2011}. This idea is challenged by low-resolution observations and simplified theoretical approaches \citep{Klimchuk_2012, Tripathi_Klimchuk_2013, Patsourakos_etal_2014} but supported by recent high-resolution observations \citep{Henriques_etal_2016, DePontieu_etal_2017b, Chintzoglou_etal_2018} and numerical modeling \citep{Martinez-Sykora_etal_2018}. \citet{Martinez-Sykora_etal_2017} performed a 2.5D radiative MHD simulation (using the Bifrost code; \citealt{Gudiksen_etal_2011}) that considered the effects of ion-neutral interactions in the chromosphere, producing spicules that match the Type-II properties mentioned above. Recently, several theoretical models addressed the problem of spicule formation, such as the 3D MHD simulation by \citet{Iijima_Yokoyama_2017} where a jet structure (matching the characteristic physical size and life-time of Type-II spicules) was produced and driven by the Lorentz force. 

There has been a controversy regarding the nature of Type-II spicules when seen on-disk, which dates back to the first observations with the \emph{Interface Region Imaging Spectrograph} (\emph{IRIS}; \citealt{DePontieu_etal_2014}): rapid brightenings along the length of the spicules suggest the upward shooting of hot chromospheric mass at plane-of-the-sky speeds as high as $\approx$300\,km s$^{-1}$ (i.e., speeds greater than the highest plane-of-the-sky speeds of Type-II spicules as seen in \ion{Ca}{2} H). Such events were termed as ``network jets'' (\citealt{Tian_etal_2014}). \citet{RouppevanderVoort_etal_2015} observed Type-II spicules on-disk and found that the Doppler velocities associated with network jets were far lower than those seen on the plane of the sky (e.g., \citealt{Tian_etal_2014}). Advanced numerical modeling also suggested that the network jets are often not a manifestation of rapid mass flows but rather rapidly moving fronts of enhanced emission produced by the rapid dissipation of electric currents (e.g., \citealt{DePontieu_etal_2017b}). In addition, unique observations from a Ly$\alpha$ rocket-borne spectroheliograph, i.e., the \emph{Very high Angular resolution ULtraviolet Telescope 2.0} (\emph{VAULT2.0}; \citealt{Vourlidas_etal_2016}), revealed a Type-II spicule in Ly$\alpha$ (plasma temperatures $\approx$10,000-15,000\,K) minutes before such network jets appeared in \ion{Si}{4} imaging ($\approx$80,000\,K) from \emph{IRIS} \citep{Chintzoglou_etal_2018}. The same work by \citep{Chintzoglou_etal_2018} revealed unambiguously the multi-thermal nature of Type-II spicules, since once the spicule appeared in transition region temperatures, the structure persisted in Ly$\alpha$ imaging, even during the moments of recurrent network jet brightenings.

The \emph{Atacama Large Millimeter/submillimeter Array} (\emph{ALMA}; \citealt{Wootten_Thompson_2009}) has recently become available to be used for the study of the chromosphere via imaging of the free-free emission (from chromospheric electrons) in millimetric (mm) wavelengths. At mm-wavelengths and under chromospheric conditions the source function, $S_\lambda$, of the free-free emission is in local thermodynamic equilibrium (LTE) and so the source function is the Planckian, $S_\lambda=B_\lambda(T)$:
\begin{equation}\label{planck1}
S_\lambda=B_\lambda=\frac{2hc^2}{\lambda^5}\frac{1}{e^{hc/(\lambda k_B T)}-1}
\end{equation}

\noindent where $k_B$, the Boltzmann's constant, $h$, the Planck's constant, $c$, the speed of light, $T$, the blackbody temperature, and $\lambda$, the wavelength of the observations. Since we are in the long-wavelength limit $hc/(\lambda k_B T)=h\nu/(k_B T) << 1$ the equation above simplifies to the Rayleigh-Jeans approximation, which is a linear relationship of the source function with the blackbody temperature. Furthermore, the mm-emission becomes optically thick over a rather narrow width of heights at any given observed line-of-sight in the corrugated chromosphere. Since the emission is optically thick we can use its brightness to determine the local plasma temperature (i.e., at the height of formation of the free-free emission). We thus define the ``brightness temperature'', $T_b$, which is the equivalent temperature a blackbody would have in order to be as bright as the brightness of the observed emission, and which is a measure of the local plasma temperature. Such measurements are enabled via interferometric imaging observations (at a frequency $\nu$) of the spectral brightness converted into $T_b$ through the Rayleigh-Jeans approximation.

\begin{equation}\label{rayleighj}
I_\nu\approx\frac{2k_B\nu^2T_b}{c^2}=\frac{2k_BT_b}{\lambda^2}.
\end{equation}

\noindent However, since the chromosphere is fine-structured and corrugated, the local conditions producing the optically-thick free-free emission can originate from quite a wide range of geometric heights; the formation height is also dependent on the electron density and thus the actual height where the free-free emission becomes optically thick is typically not well known \citep{Carlsson_Stein_2002, Wedemeyer_etal_2007, Loukitcheva_etal_2015, Martinez-Sykora_etal_2020b}. \citet{Rutten_2017} predicted that \emph{ALMA} will observe fibrils along the canopy which are optically thick and thus will mask emission from lower heights. We further investigate the \emph{ALMA} mm-emission formation height problem in a companion publication (Chintzoglou et al 2020b; ``Paper II''). 

An interesting application of \emph{ALMA} observations for the study of spicules was presented in the works of \citet{Yokoyama_etal_2018} and \citet{Shimojo_etal_2020}, in which they focused on spicules seen at the limb (the latter, however, captured a macro-spicule). These studies faced challenges due to low signal-to-noise ratio in the \emph{ALMA} limb observations (primarily due to an interferometric ``knife-edge'' effect when observing at the limb; \citealt{Shimojo_etal_2017}), also worsened by  confusion/overlapping from foreground/background structures along the line-of-sight (hereafter, LOS),  the latter being typical for observations of spicules at the limb.

In this paper, we composed and analyzed a unique and comprehensive dataset from joint observations with \emph{ALMA}, \emph{IRIS}, and the \emph{Solar Dynamics Observatory} (\emph{SDO}; \citealt{Pesnell_etal_2012}). Our dataset is most appropriate for investigating the rich dynamics of the solar chromosphere and transition region in plage and its peripheral areas. In particular this dataset is excellent for the study of spicules thanks to the synergy of high spatial and temporal resolution spectral and imaging observations by \emph{IRIS} with high time-cadence and unique temperature diagnostic capabilities from \emph{ALMA} interferometric observations. 

This paper is organized as follows: in \S~\ref{obs} we provide a description of the observations used in this work. In \S~\ref{analysis} we present the analysis of the observations and the synthetic observables, and in \S~\ref{results} our results, followed by a summary and conclusions in \S~\ref{conclusion}.

\section{Observations}\label{obs}

\begin{figure}
	\centering
	\includegraphics[width=5.8in]{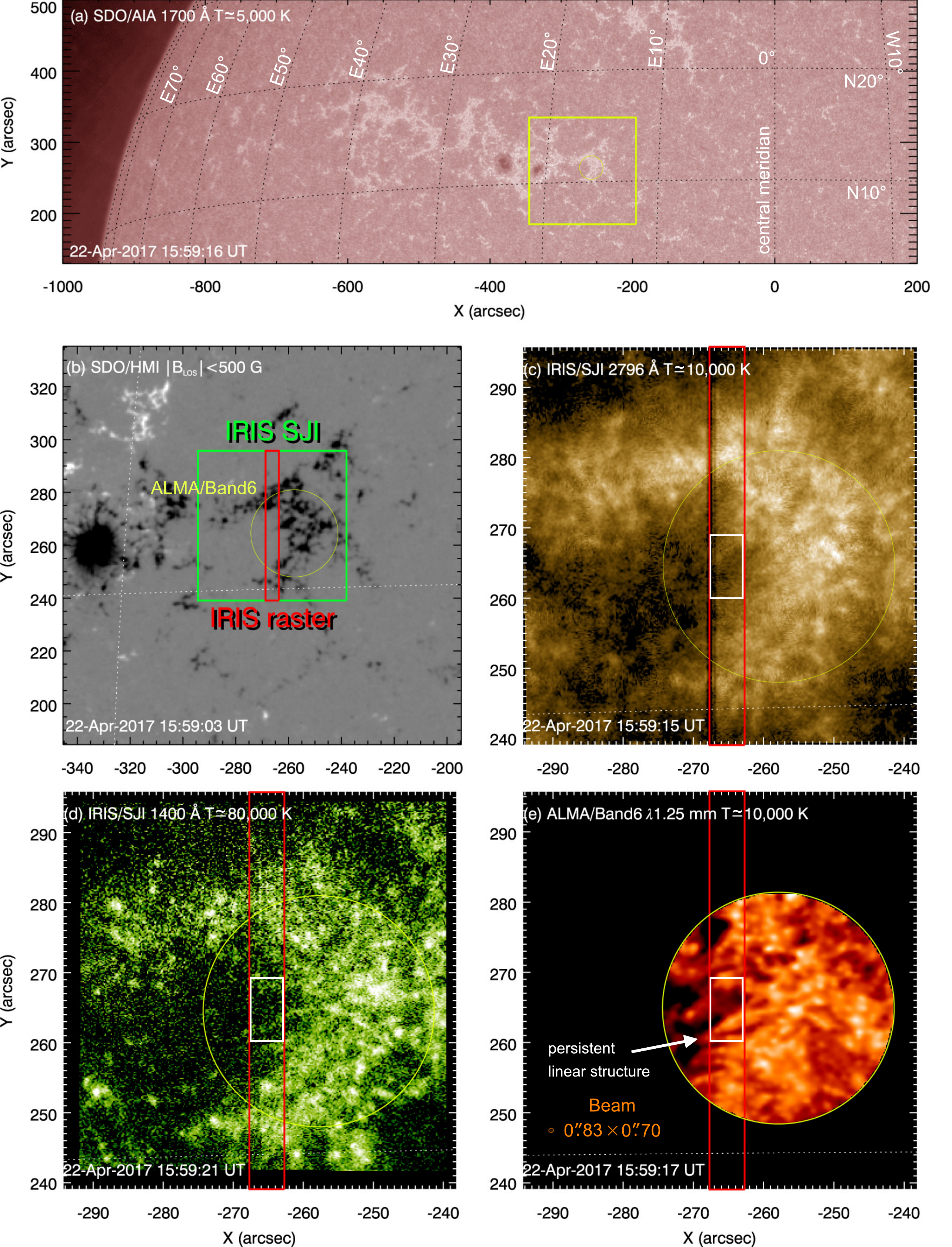}

	\caption{ (a) Context figure from \emph{SDO}/AIA 1700\,\AA\ showing the location of the plage region observed by \emph{IRIS} and \emph{ALMA}. (b) The magnetic field distribution from \emph{SDO}/HMI is shown for the boxed area in panel (a). The same panel contains a narrow width inset plot of the \emph{IRIS}/SJI 2832\,\AA\ (\ion{Mg}{2} continuum). (c) \ion{Mg}{2} 2796\,\AA\ and (d) \ion{Si}{4} 1400\,\AA\ corresponding to the \emph{IRIS}/SJI FOV (green box) in panel (b) . The corresponding \emph{ALMA}/Band6 FOV is shown in panel (e). The white rectangle marks the FOV of the raster cutout shown in Figure~\ref{FIG_2} (i.e., trimmed along the y-direction to focus on the spicule. Note the persistent linear-like structure near the future location of the spicule).
	}\label{FIG_1}
\end{figure}

Our observations targeted a plage region in the leading part of NOAA AR12651 on 22-Apr-2017, centered at heliographic coordinates N11$\degr$E17$\degr$, or at ($x$,$y$)=(-260$\arcsec$, 265$\arcsec$) in helioprojective coordinates (Figure~\ref{FIG_1}a). The polarity of the photospheric magnetic field in that plage region was negative, i.e., of the same sign as the leading sunspot of NOAA AR12651 (Figure~\ref{FIG_1}b). The overall spatial distribution of the plage fields in the target appeared semicircular in shape, as organized around the outer boundary of a supergranule -- the latter being evident by the very low magnetic flux density in the core of the supergranule's area (Figure~\ref{FIG_1}b). The common \emph{IRIS} and \emph{ALMA} field of view (FOV) contained part of that plage, including a photospheric pore, and also intersected the supergranular cell center, which appears in the chromosphere as a region of low background intensity (e.g., Figure~\ref{FIG_1}c). The latter allowed us to study morphological structures, such as fibrils or loops, resolved in high contrast due to the weak background emission at the supergranular cell center.

\begin{figure}
	\centering
	\includegraphics[width=\linewidth]{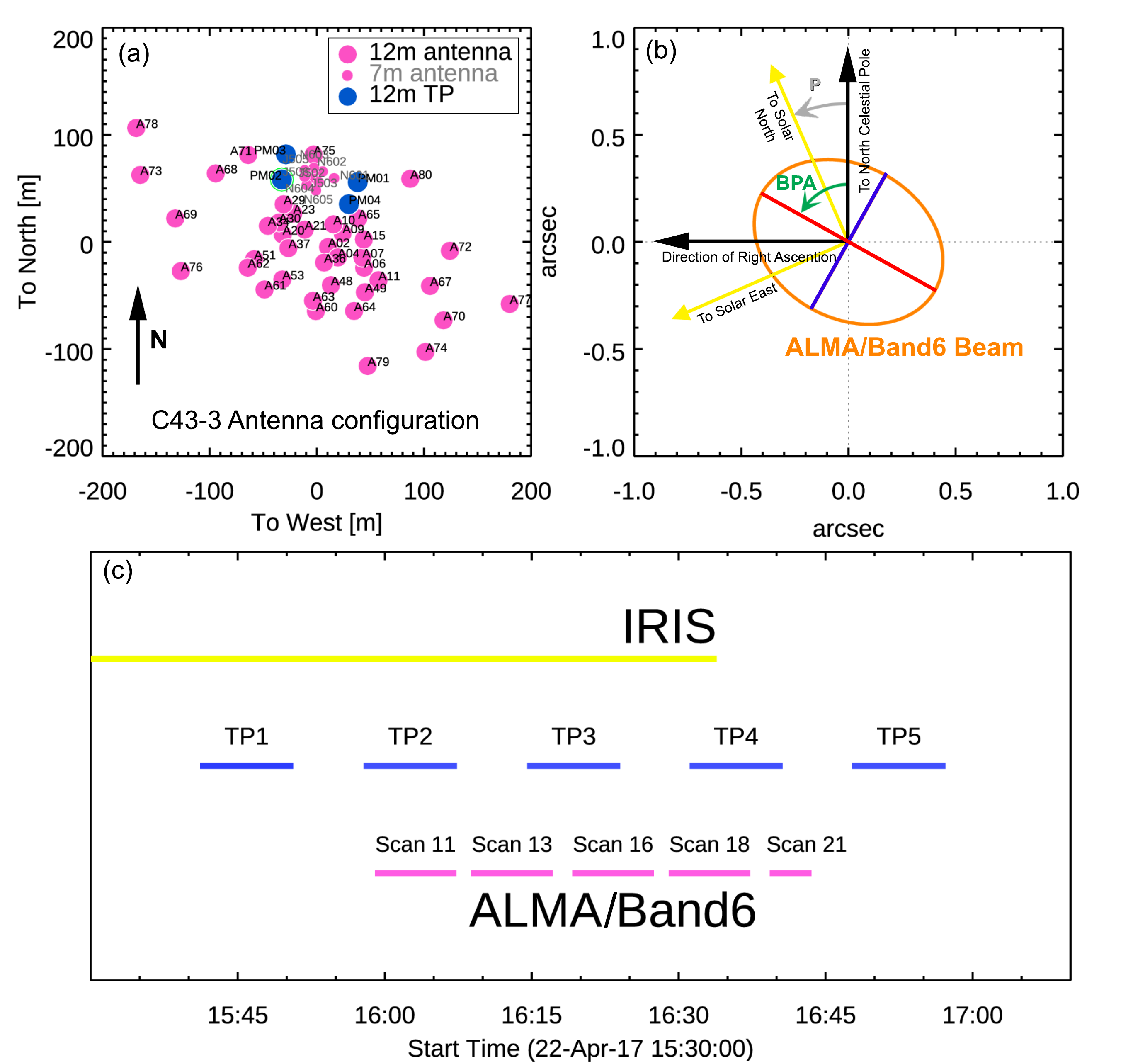}
	
	\caption{ (a) \emph{ALMA} antenna configuration at the time of the observations. Three out of forty-three 12\,m-array antennas were excluded due to problems with their calibration; similarly, only nine antennas in the 7\,m array were included. For the TP measurements only one 12\,m TP antennas was used (highlighted with a green circle). The labels correspond the specific identification name of each antenna. (b) The resolution element in the interferometric imaging (i.e., the ``beam''), is a 2D gaussian full-width at half-maximum with a major (red) and a minor (blue) axis. The beam's position angle (BPA) is measured between the north celestial pole and the beam major axis counter-clock-wise, like the solar P angle. Note that the solar P angle at the time of the observations was negative, i.e., $\approx$-25$\degr$. (c) Timeline of joint \emph{ALMA}/Band6 and \emph{IRIS} observations. Scans are interrupted by special scans (not shown) required for calibrations of the interferometric array. 
	}\label{FIG_TIMELINE}
\end{figure}

\emph{ALMA} is a general-purpose ground-based telescope located at an elevation of 5,\,000\,m in the Atacama desert in Chile, operating in wavelengths ranging from 0.32-3.57\,mm, or frequencies from 84-950\,GHz. It includes two arrays of antennas designed to perform Fourier synthesis imaging together or separately: 1) one array is composed of fifty 12\,m antennas that can be moved to separations as large as 16\,km; 2) the other array, the Atacama Compact Array (ACA), is a fixed array of twelve 7\,m antennas designed for interferometry plus four 12m total power (TP) antennas (see below). \emph{ALMA} was commissioned for solar observing in 2014-2015 and was first made available to the community for scientific observations in 2016 (see \citealt{Wedemeyer_etal_2016, Shimojo_etal_2017, White_etal_2017}).  Solar observations use both the fifty-antenna 12\,m array and the ACA 7\,m array as a single array. The fifty-antenna array was only available in the most compact \emph{ALMA} antenna configurations (C43-1, C43-2, and C43-3) in 2016-2017 \citep{Shimojo_etal_2017}. In addition, pending ongoing commissioning activities, solar observations were initially only available at 3\,mm (100\,GHz; Band3) and at 1.25\,mm (239\,GHz; Band6). The observations reported here used the \emph{ALMA} C43-3 antenna configuration which provides antenna separations ranging from 14.6\,m to 500\,m. The ACA 7\,m antennas provide antenna spacings ranging from 8.7\,m to 45\,m. They were used to image the plage region with an angular resolution of $\approx 0\farcs8\times 0\farcs$7 as determined by the synthesized beam. The array configuration and the beam aspect are summarized in Figure~\ref{FIG_TIMELINE}. It is important to note, however, that any interferometric array acts as a spatial filter. It does not measure spatial frequencies smaller than the minimum antenna separation in the array, corresponding to the largest angular scales in the source. For the Sun, most of the power is on the largest angular scales and it is therefore important to recover them if photometry is required for the science goals. The 12\,m TP antennas provide this information by mapping the full disk of the Sun with an angular resolution corresponding to that of a 12\,m antenna (note that usually, and in our observations, only one TP antenna was used). Roughly speaking, TP maps provide measurement on angular scales $>24\arcsec$ and the joint 12\,m/ACA array provides measurements on angular scales $<24\arcsec$ for the Band6 observations reported here. The two types of measurements are combined in post-processing using the ``feathering'' technique (e.g., \citealt{Cotton_2017}).

\emph{IRIS} is a space-borne imaging spectrograph operating in near-ultraviolet (NUV) and far-ultraviolet (FUV) wavelengths, designed to improve our understanding of the solar chromosphere and transition region. It produces two types of observations: (1) ``slit-jaw'' context imaging (hereafter SJI) in pass bands centered at \ion{Mg}{2} k 2796\,\AA, \ion{C}{2} 1330\,\AA, \ion{Si}{4} 1400\,\AA, and Mg continuum 2832\,\AA; and (2) spectral raster scans (in short, rasters), the latter covering passbands containing several chromospheric and transition region lines found within a NUV and FUV spectral window (range 2783-2834\,\AA\ for NUV and two ranges, 1332-1358\,\AA\ and 1389-1407\,\AA\ for FUV). The rasters are produced at a variety of slit sampling positions, giving the choice of dense (0$\farcs$35 steps), sparse (1$\arcsec$), or coarse (2$\arcsec$) rasters in lines like \ion{Mg}{2}, \ion{Si}{4}, \ion{C}{2}, etc. \emph{IRIS} provided coordinated observations with \emph{ALMA}, starting on 2017-04-22 13:29:36 UT until 16:33:53 UT. The \emph{IRIS} observing mode for this dataset was OBS 3620502035, producing SJI images at the aforementioned passbands at 0$\farcs$16 pix$^{-1}$ with FOV 60$\arcsec\times$65$\arcsec$ at 13\,s cadence and also 16-step dense rasters (i.e., 0$\farcs$35 step in helioprojective $x$ direction and 0$\farcs$16 in $y$ across the slit) covering 5$\arcsec\times$60$\arcsec$ at 26\,s cadence of an area intersecting plage, a photospheric pore, and the core of the supergranule (e.g., red rectangle in Figure~\ref{FIG_1}bcd). The spectral resolution in the rasters was 0.0256\,\AA\ for the NUV window and 0.053\,\AA\ for the FUV. Since the exposure time was 0.5\,s in order to optimize raster cadence for \ion{Mg}{2} k spectra, the 1400\,\AA\ SJI images show low signal and have been summed, using a 3-frame (``boxcar'') temporal averaging on the image series. Such averaging improves the signal-to-noise ratio (S/N) enough so as to resolve dynamic features (at an effective timescale of 39\,s). Unless otherwise specified in this paper, we do not perform temporal averaging on the raster data since they are of higher S/N.

\emph{ALMA} observed in Band6 from 15:59 UT to 16:38 UT and produced 5 scans of the target region, owing to four breaks for interferometric calibration (spanning from 1.75 to 2.25 minutes; Figure~\ref{FIG_TIMELINE}c). In addition, \emph{ALMA} observed in Band3 between 17:20 UT to 17:56 UT but without \emph{IRIS} support. Thus, the overlapping time range between \emph{ALMA} and \emph{IRIS} amounts to a total of 34 mins of Band6 data (Figure~\ref{FIG_TIMELINE}; Band3 not shown).
Within this period of time, \emph{ALMA} was able to capture in ultra-high cadence (2\,s) rich dynamic activity and interesting evolution of linear-like structures, including indications for shocks in the region just above the plage \bf(we address this in a separate publication, Paper II: Chintzoglou et al 2020b)\rm. In this work, we focus on a slender and dynamically evolving linear-like structure (\S~\ref{results}), resolved with \emph{ALMA}/Band6 observations at high cadence and spatial resolution.



\section{Analysis}\label{analysis}

\subsection{Reduction of \emph{ALMA} and \emph{IRIS} observations}\label{Reduction}

The calibrated \emph{ALMA} data was obtained from the \emph{ALMA} Science Archive and processed with the Solar \emph{ALMA} Pipeline (SoAP, Szydlarski et al. in prep., see also \citealt{Wedemeyer_etal_2020} for details). 
SoAP is developed by the SolarALMA project in Oslo in collaboration with the international solar \emph{ALMA} development team and is based on the Common Astronomy Software Applications (CASA) package \citep{McMullin_etal_2007}. 

Imaging with SoAP uses the multi-scale (multi-frequency) CLEAN algorithm \citep{Rau_Cornwell_2011} as implemented in CASA, self-calibration for a short time window of 14\,s, primary beam correction, and combination of the interferometric data with the TP maps via the ``feathering’’ method. 
For the Band6 observations reported here, the TP maps were scaled to a mean quiet Sun brightness of 5,900 K as recommended by \citet{White_etal_2017}, who quote a nominal uncertainty of 5\%. 
For the interferometric part, all frequency channels are used to produce one continuum image for each time step at 2\,s cadence. 
The final result is a time sequence of absolute brightness temperature continuum maps at 2\,s cadence with short ($\sim$2\,min) calibration breaks that divide the sequence into 10\,min segments or ``scans''.

Ensuring precise co-registration of the feathered \emph{ALMA}/Band6 interferometric maps with the \emph{IRIS} rasters is imperative for the successful analysis of this comprehensive dataset. For this purpose, we used imaging observations in 1700\,\AA\ from the Atmospheric Imaging Assembly (AIA; \citealt{Lemen_etal_2011}) on board \emph{SDO} to coalign \emph{IRIS}/SJI 1400\,\AA, 2796\,\AA\ and the FUV/NUV raster image series with other \emph{SDO} observations, such as line-of-sight magnetograms (hereafter, BLOS) produced with the Helioseismic Magnetic Imager (HMI; \citealt{Scherrer_etal_2012}). To coalign \emph{ALMA} with \emph{IRIS}, we exploited the very high degree of similarity between morphological structures seen in \emph{ALMA}/Band6 maps and \emph{IRIS}/SJI 1400\,\AA\ images. This can be readily seen in Figure~\ref{FIG_1}, e.g., by comparing the bottom panels (d) and (e) for bright structures in the plage region. \bf We address the origin of this interesting similarity in Paper II. \rm

\subsection{General Morphology of the Structure Seen in \emph{IRIS} and \emph{ALMA}/Band6 Observations}

\begin{figure*}
	\includegraphics[width=\linewidth]{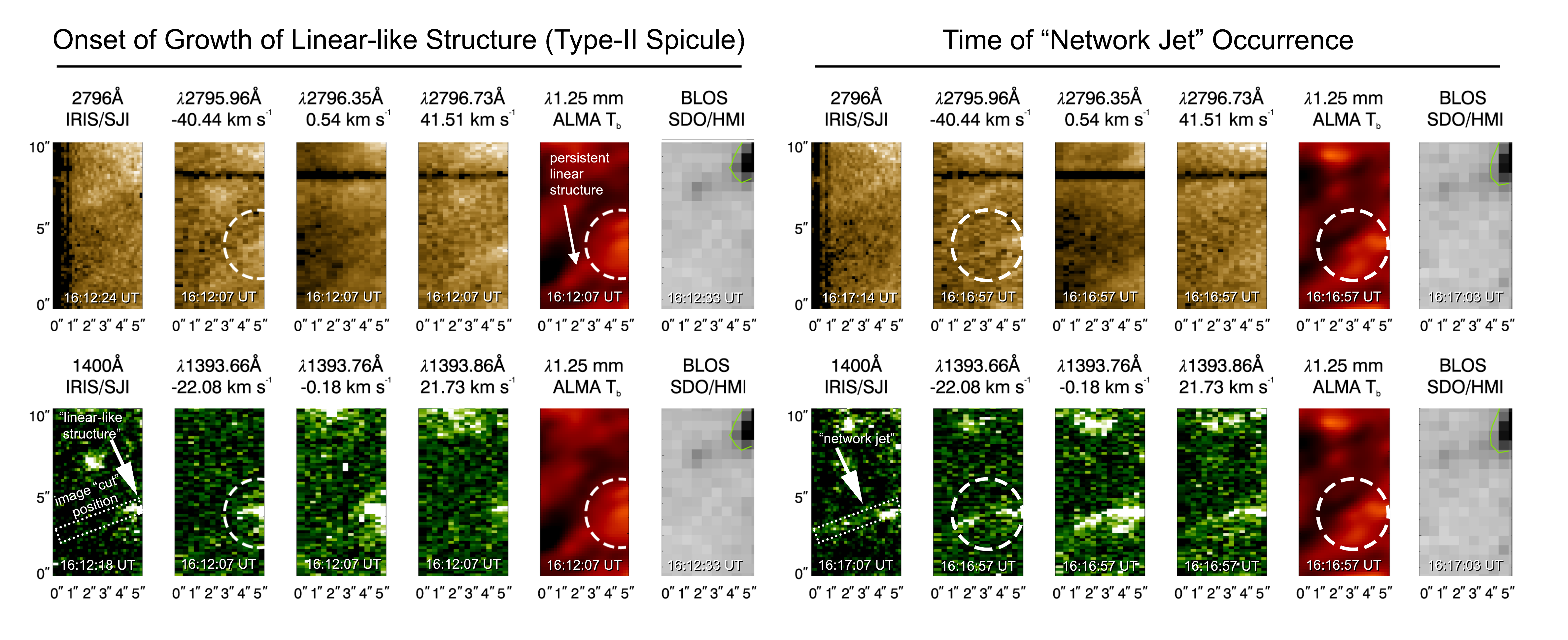}
	\caption{Inter-comparison of \emph{IRIS} and \emph{ALMA}/Band6 observations at two different times in the evolution of the linear-like structure. Raster scans are shown at selected wavelength positions in \ion{Mg}{2} (top) and \ion{Si}{4} (bottom) showing the clear appearance of the rapidly evolving structure in \ion{Mg}{2} and \ion{Si}{4} (dashed circles). In the 1400\,\AA\ \emph{IRIS}/SJI panels we show the position of the image cut we performed on each observable to produce Figure~\ref{FIG_3}. In the panels with the \emph{SDO}/HMI magnetograms (scaling clipped at $\pm$250\,G) we overplot an isocontour of $\pm$100\,G.}\label{FIG_2}
\end{figure*}

The \emph{IRIS}-\emph{ALMA} FOV captured several slender linear-like structures; some of them appeared as persistent and slowly evolving while others showed dynamic behavior over the duration of the observations. In Figure~\ref{FIG_2} we present a dynamic and prominent linear structure when it first appeared at around 22-April-2017 16:12 UT in the western part of the \emph{IRIS} raster FOV (e.g., see the left group of panels in Figure~\ref{FIG_2}; structure in the \emph{IRIS}/SJI 1400\,\AA\ panel pointed by an arrow; for a movie see electronic version of Figure 1 in Paper II). Over the minutes that followed, this structure grew from west to east (e.g., see right group of panels of Figure~\ref{FIG_2} around 16:17 UT; structure pointed by an arrow). We focus our study on the aforementioned dynamic linear structure in Figure~\ref{FIG_2} since: (1) its outstanding dynamic nature is interesting; and because (2) for the most part of its time-evolution it is observed against low background emission (due to the supergranular core's weak magnetic fields; \emph{SDO}/HMI panels in Figure~\ref{FIG_1} and~\ref{FIG_2}), with the latter allowing us to take measurements that are free of confusion from background structures. 

Previous studies on the on-disk counterparts of Type-II spicules identified them through signatures like short-lived asymmetries of chromospheric/transition region spectral lines, which render the lines skewed to the blue or red wing. This effect is described as a ``Rapid Blue-shifted (or Red-shifted) Excursion'' (RBEs and RREs; e.g., \citealt{Langangen_etal_2008, RouppevanderVoort_etal_2015}). Here, by exploring the spectral information in the \emph{IRIS} raster series at the corresponding time in the SJI maps of 16:12 UT in Figure~\ref{FIG_2}, we can see that the structure appears faintly in raster images at the blue wing of the \ion{Mg}{2} k line, corresponding to Doppler blueshifts of -40\,km s$^{-1}$ and also in the \ion{Si}{4} 1393\,\AA\ line, here seen clearly at -22\,km s$^{-1}$ (see rasters in left group of panels of Figure~\ref{FIG_2}; dashed semicircle in \ion{Mg}{2} k and \ion{Si}{4} rasters, and in \emph{ALMA}/Band6 common FOV). The structure is not clearly seen in line positions which correspond to Doppler redshifts as high as the blueshifts (see Figure~\ref{FIG_RBE} for $\lambda-t$ plots sampling the structure at three different positions across its length). While the evidence is clear in \ion{Si}{4}, \ion{Mg}{2} also shows such signature at least in ``position B'', albeit with low S/N. As a result we conclude that this feature is characterized as an RBE event. The fact that such slender structure exhibits the spectral signatures of an RBE event further suggests that the linear-like structure is the on-disk manifestation of a Type-II spicule shooting mass upwards (blueshifts of $\approx$-50\,km s$^{-1}$ are typical for RBEs associated with Type-II spicules;  \citealt{RouppevanderVoort_etal_2015, DePontieu_etal_2017b}). Later on, the structure grows further to the east and brightens along its length in \ion{Si}{4} (right group of panels of Figure~\ref{FIG_2} at 16:17 UT). Throughout the evolution of the linear structure, in the \emph{ALMA}/Band6 data the latter appears somewhat wider and shorter in length as compared to its appearance in the blue wing of \ion{Mg}{2} k and in \ion{Si}{4}. However, considering that \emph{ALMA}/Band6 and \emph{IRIS}/\ion{Mg}{2} k observations probe similar ranges of temperatures (e.g., \citealt{Bastian_etal_2017}), such differences in morphological structures between \ion{Mg}{2} k and \emph{ALMA}/Band6 could be also owing to the difference in spatial resolution between \emph{IRIS} data (0$\farcs$35 pixel$^{-1}$ in $x$ and 0$\farcs$16 pixel$^{-1}$ in $y$) and in \emph{ALMA}/Band6 (here, beam size $\approx$0$\farcs$7 at best). Another possibility, despite the common ranges in temperatures in both observables, is that we may not be looking at the same part of the structures as the observed intensities may be coming from different parts of the same multithermal event \bf (we  address this in Paper II)\rm. In \ion{Si}{4} rasters, different parts of the structure can be seen clearly at -22\,km s$^{-1}$ and at +22\,km s$^{-1}$ away from line center, although the signal in those line positions is lower than at the line core. 

\subsection{Dynamic Evolution of the Type-II Spicule}

In order to properly study the dynamics, we proceed by making an image ``cut'' in the raster series aligned along the principal axis of the spicule (i.e., its longest spatial dimension; position of ``cut'' shown in Figure~\ref{FIG_2} 1400\,\AA\ SJI panels only for reference as here we focus on the raster series) and produce a space-time plot (hereafter referred to as ``$x-t$ plot''). For our \emph{ALMA}/Band6 observations we make an $x-t$ plot of the brightness temperature, $T_b$. For \emph{IRIS} \ion{Mg}{2} k and \ion{Si}{4} we produce $x-t$ plots in selected wavelength positions sampling the spectral lines around their rest positions, and we stack these $x-t$ plots sorted in the velocity space as shown in Figure~\ref{FIG_3} (red line marks the time of frames in Figure~\ref{FIG_2}). In the same figure, we also show \emph{IRIS} \ion{Mg}{2} k and \ion{Si}{4} $x-t$ plots after integrating the rasters in wavelength (within 0.7\,\AA\ and 0.2\,\AA\ from line rest positions, respectively). We highlight the parabolic $x-t$ profile of the spicule with yellow dotted lines in Figure~\ref{FIG_3}. The parabola is extracted from the \ion{SI}{4} wavelength-integrated $x-t$ envelope and replicated and overplotted in selected panels of the other observables to serve as a guide for comparisons. The integrated \ion{Si}{4} shows a parabolic profile in a more complete way. In comparison, the profiles in specific wavelength positions appear partial, albeit consistent with Dopplershift modulation due to ascending and descending plasma motions along the evolution timeline. On the other hand, the integrated \ion{Mg}{2} $x-t$ profile shows a less clear picture, as we discuss below. \ion{Mg}{2} is a complex spectral line that is typically optically thick and formed under non-LTE conditions. Disk counterparts of spicules appear as features that can be both brighter or darker than neighboring features during their complex temporal evolution (e.g., \citealt{RouppevanderVoort_etal_2015, Bose_etal_2019}). This complex evolution is a key reason for the lack of clear parabolic evolution compared to that seen in \ion{Si}{4}. The structure initially grows from west to east in both the \emph{ALMA}/Band6 and \emph{IRIS} raster plots of Figure~\ref{FIG_2}. 

At a first glance, the integrated intensities between the \ion{Si}{4} and \ion{Mg}{2} k $x-t$ plots appear anti-correlated in these structures: \ion{Si}{4} seems to be brightest where \ion{Mg}{2} is darkest (dotted circles in Figure 5; compare the intensity in the $x-t$ plots between times 16:15-16:20\,UT; there seems to be an intensity depression in \ion{Mg}{2} $x-t$). This can be understood as cool material is injected in the corona (i.e., the spicule) with its bulk appearing in absorption in \ion{Mg}{2} and its interface with the corona seen in emission in transition region temperatures (i.e., \ion{Si}{4}). \bf We present this interpretation in detail by carefully investigating the similarities and anti-correlations of intensities between different observables in Paper II\rm. In order to interpret this behavior properly, we compare our observations to our synthetic observations (see next section). First of all, these three observables represent very different properties of the plasma, i.e., \emph{ALMA}/Band6 provides brightness temperature at an unspecified height, \ion{Mg}{2} intensity comes from radiation that is optically thick (chromosphere), and \ion{Si}{4} intensity comes from radiation that is optically thin (transition region). It is also interesting to note that \ion{Mg}{2} k does not reach the parabola as close as \emph{ALMA}/Band6 does. \ion{Si}{4}, on the other hand, is contained under the parabola rather well, also hosting transient brightenings (bright areas associated with the network jet). 

However, overall, the behavior seen between these observables is quite similar -- for instance, compare the parabolic profiles in the $x-t$ plots of \ion{Mg}{2} rest $\pm$27\,km s$^{-1}$ and \emph{ALMA}/Band6, where the trace of the \emph{ALMA}/Band6 emission seems to be co-temporal to similar intensity enhancements in \ion{Mg}{2} k (i.e., between 16:12-16:14\,UT), but not beyond that, where \ion{Mg}{2} k does not seem to extend as high as \emph{ALMA}/Band6. We also note that at \ion{Mg}{2} rest -55 \,km s$^{-1}$ the profile in $x-t$ is more pronounced than that at +55 \,km s$^{-1}$, which is a manifestation of the spicule's RBE-nature. In fact, the \ion{Mg}{2} rest -55 \,km s$^{-1}$ profile's duration and elongation corresponds relatively well to the $x-t$ profile in \emph{ALMA}/Band6 up until $\sim$16:19 UT (the gaps in the \emph{ALMA}/Band6 observations were due to breaks for calibration purposes).

The \ion{Si}{4} $x-t$ plots (Figure~\ref{FIG_3}; green-colored plots) suggest upflows in the spicule during its ascent (or ``growth phase''), which is then followed by downflows (the ``receding phase''). The ascending part of the parabolic profile is quite localized (appears as a bright thin trace) and is seen clearly in the blue wing (even seen at a blueshift as high as -66\,km s$^{-1}$). At around 16:16-16:17 UT, a network jet brightening occurs in the spicule (dotted circle in Figure~\ref{FIG_3}) with an apparent (i.e., projected on the plane of the sky) speed of $\approx$95 km s$^{-1}$. Since the time cadence of the rasters was at 26\,s this apparent speed is only a lower limit. We note that the network jet brightening started just after the $x-t$ profile's ``apex'' point and it is best seen in redshifts, which is consistent with mass motions moving away from the observer. An alternative explanation could be that the spicular column is curved with respect to our line of sight. We should also note, however, that \ion{Si}{4} integrated intensity appears to come just above the instantaneous maximum extent of \emph{ALMA}/Band6, at least before the onset of the network jet brightening (where we do not have \emph{ALMA}/Band6 data). We further illustrate this fine point in the following sections.


\subsection{Bifrost Simulation of Type-II Spicules and Synthesis of \emph{ALMA} and \emph{IRIS} Observables}\label{observables}

In order to further investigate the nature of the Type-II spicule, we employ a 2.5D MHD numerical model based on the Bifrost code \citep{Gudiksen_etal_2011}, covering the upper solar convection zone all the way to the low corona (extending up to 40\,Mm from photosphere). This model includes the effects of non-equilibrium (in short, NEQ) ionization (for hydrogen and helium; \citealt{Leenaarts_etal_2007, Golding_etal_2016}) and ambipolar diffusion \citep{Martinez-Sykora_etal_2017,Nobrega-Siverio_etal_2020}. \citet{Martinez-Sykora_etal_2020a} compared this model with an equivalent model without NEQ ionization in LTE conditions. The most important difference for the matter of interest of this paper is that the NEQ ionization increases the electron density, $N_e$, in the upper-chromosphere due to the large recombination time-scale. In addition, any heating (for instance, due to ambipolar diffusion) or cooling (i.e., due to adiabatic expansion) will increase or decrease the temperature instead of ionizing or recombining the plasma owing to large ionization/recombination time-scales producing multi-thermal structures, such as Type-II spicules or low-lying loops. As we discuss later on, this improves the agreement between synthetic observables and the observations (e.g., as compared to \citealt{Martinez-Sykora_etal_2017}). The spatial scale in the simulation was 14\,km (grid point)$^{-1}$. We stored the output of the model at a cadence equivalent to 10\,s of solar time. After the simulation relaxed from the initial condition (which took $\approx$50\,mins of solar time), the remaining total duration of the simulation we analyze here represents $\approx$10\,min of solar time. Using this 2.5D model output we synthesize observables from the physical conditions in the Bifrost model (for each snapshot of the simulation) that correspond to the observed chromospheric emission from a vantage point overlooking the simulation from the top of the domain, essentially simulating ``sit-and-stare'' slit observations near solar disk center.

The simulation produces Type-II spicules in several locations in the computational domain, in-between regions of emerging flux and plage (the latter containing dynamic fibrils). In order to perform a comparison of the physics and the evolution of the observed spicule with those in the simulation, we focus on a particular region, i.e., at $x=[40,45]$\,Mm, where two neighboring spicules (hereafter ``spicule 1'' and ``spicule 2'') are seen to develop at a favorable angle with the LOS (e.g., Figure~\ref{FIG_4}ab, annotated and pointed with arrows), which results in well-isolated parabolic profiles in the various $x-t$ plots of synthetic observables presented in Figure~\ref{FIG_4}c-h. We assume that the spicule’s orientation is not such that the LOS intersects it perpendicularly over its length, as the latter seems an extreme case for its orientation (likewise for the case where the spicule is viewed along its axis). Thus, the geometry in the model seems reasonable for the interpretation of the observations.”

We compute synthetic \emph{ALMA}/Band6 observations from our simulations at a single wavelength position of  $\lambda$=1.2\,mm. For this, we used the LTE module of the Stockholm Inversion Code (STiC code; \citealt{delaCruzRodriguez_2016,delaCruzRodriguez_2019}). 
It computes the partial densities of all species considered in the calculations using the electron densities and gas pressure stratifications from the simulation. For the construction of the equation of state (EOS) we use the first 28 elements of the periodic table with 3 ionization stages, except for H. For H we used a simplified EOS that only includes H$_2$ molecules. Many of those elements will not contribute at mm-wavelengths, but they are included in the opacity and the background opacity (for the latter we consider the partial densities only for elements that are major contributors). Continuum opacities are calculated using routines ported from the ATLAS code \citep{Kurucz_1970}, which include the main opacity source at mm-wavelengths (free-free hydrogen absorption; e.g., see
\citealt{Wedemeyer_etal_2016}). We also note that free-free and bound-free opacity processes from H and H$^{-}$ are also included in addition to those for the other atoms considered. The emergent intensity
is calculated using a formal solver of the unpolarized radiative
transfer equation based on cubic-Bezier splines
\citep{Auer_2003,delaCruzRodriguez_2013}. The viewing geometry chosen was for an ``observer'' looking top-down the domain (i.e., assuming a LOS along the vertical direction in the simulation). We show the \emph{ALMA}/Band6 $x-t$ plot in Figure~\ref{FIG_4}c.

 In order to synthesize the \ion{Mg}{2} k line emission from the simulated spicules we used the RH radiative transfer code (\citealt{Uitenbroek_2001, Pereira_Uitenbroek_2015}). The synthesis was done individually for each vertical column of the simulation domain, for an observer looking ``top-down'' the simulated box (i.e., the z-axis in Figure~\ref{FIG_4}ab). We show an $x-t$ plot in Figure~\ref{FIG_4}e, after integrating \ion{Mg}{2} k in a wavelength range of $\Delta\lambda=0.7$\,\AA\ centered at the k3 rest wavelength at 2796.35\,\AA\ 
 
Lastly, we computed the \ion{Si}{4} 1393\,\AA\ spectrum assuming ionization equilibrium and also assuming optically-thin emission. From the spectrum we compute the intensity by integrating the locally-deterimined emissivity along the same LOS as before. We thus produced full spectra, which can be used to compare to  the \emph{IRIS} \ion{Si}{4} 1393\,\AA\ rasters. An illustration of the \ion{Si}{4} synthesis is provided as an $x-t$ plot in Figure~\ref{FIG_4}g (in this example, we provide the total intensity of \ion{Si}{4} 1393\,\AA).

Furthermore, for the optically-thick synthetic observables (i.e, \ion{Mg}{2} k, \emph{ALMA}/Band6), we use the height where the optical depth $\tau=1$ as a function of $\lambda$ to interpret the diagnostic information we get from the \emph{IRIS} \ion{Mg}{2} k and \emph{ALMA}/Band6 optically-thick observations. This quantity is commonly referred to as the \emph{formation height} of a line. For demonstration purposes, here we show the maximum $\tau=1$ height that the \ion{Mg}{2} k has formed at for the wavelength range $\Delta\lambda=0.7$\,\AA\ from its rest wavelength position (Figure~\ref{FIG_4}f). Note that in our discussion we make use of the $\tau=1$ height also at specific wavelength positions (more in \S~\ref{results}). 
The \ion{Mg}{2} k line typically includes three components: k2v and k2r form in the wings and originate from the low chromosphere whereas k3, often in absorption, forms in the upper chromosphere.
Conversely, Figure~\ref{FIG_4}f reflects that the integrated \ion{Mg}{2} k probes the various structures at a range of geometric heights below that maximum $\tau=1$ height (i.e., it is an upper limit), while \emph{ALMA}/Band6 roughly observes the spicular plasma at consistently greater heights than the wavelength-integrated \ion{Mg}{2} k observable (more in \S~\ref{results}). 

The optically-thin \ion{Si}{4} synthetic observable captures the basic qualitative behavior of the $x-t$ plot from the observations (Figure~\ref{FIG_3}; green-colored plot), in that it tends to highlight the parabolic profile of spicules 1 and 2. We expand on the similarities seen between the synthetic observables and the observations in the next section \S~\ref{results}. Intensities in \emph{ALMA}/Band6 and \ion{Mg}{2} k seem to almost always originate from below that \ion{Si}{4} ``envelope'', with \emph{ALMA}/Band6 intensities extending closer to that \ion{Si}{4} envelope than \ion{Mg}{2} k. These diagnostics are formed in a region with very strong gradients and very rapid temporal evolution. This means that even though they may be formed in close proximity, they can still be substantially different. To illustrate this point better, we calculate the geometric height of maximum emission for \ion{Si}{4} and we show the resulting $x-t$ plot in Figure~\ref{FIG_4}h. As for the \ion{Mg}{2} k line, it forms over a broad range of heights -- \bf we discuss its diagnostic information in comparison to the other observables in Paper II.\rm 

In Figure~\ref{FIG_5}b we present the \emph{ALMA}/Band6 source function, $S_\nu$, calculated separately at each column in the simulated domain, and in Figure~\ref{FIG_5}a the resulting contribution function, $g_\nu$,

\begin{equation}
g_\nu = S_\nu e^{-\tau_\nu} \alpha_\nu = g_\nu(N_e, T) 
\end{equation}

\noindent where $\alpha_\nu$, the monochromatic absorption coefficient, and $\tau_\nu$ the optical depth for $\nu=240$\,GHz ($\lambda=1.25$\,mm). The source function shows at each column in the domain the range of geometric heights where the (optically-thick) \emph{ALMA}/Band6 free-free emission is forming. The synthetic \emph{ALMA}/Band6 spectral intensity, $I_\nu$, is obtained by integrating the contribution function over the geometric heights along the LOS, $z$, as

\begin{equation}\label{spectralint}
I_\nu = \int g_\nu dz = \int S_\nu e^{-\tau_\nu} \alpha_\nu dz .
\end{equation}

\noindent Pairing the above eq.~(\ref{spectralint}) with the Rayleigh-Jeans approximation of eq.~(\ref{rayleighj}) we get a map for brightness temperatures, $T_b$. For comparison, in Figure~\ref{FIG_5}d we show a map of the total (i.e., wavelength-integrated) emissivity for \ion{Si}{4} 1393\,\AA, $\eta=\int \eta_\nu c/\lambda^2 d\lambda$, with the geometric height where \emph{ALMA}/Band6 $\tau=1$. This shows that while the total intensity from \ion{Si}{4},

\begin{equation}
I = \int \eta dz  
\end{equation}

\noindent (i.e., in the $x-t$ plot of Figure~\ref{FIG_4}g) may be coming from different geometric heights in the solar atmosphere along the LOS (being optically-thin) -- but, nevertheless, delineating the spicules -- the \emph{ALMA}/Band6 optically-thick emission comes from a similarly corrugated $\tau=1$ height (but locally within a small width from that $\tau=1$ line; Figure~\ref{FIG_5}ab) following closely the height variation of \ion{Si}{4} total emissivity (panel d). Figure~\ref{FIG_5}cd demonstrates that \emph{ALMA}/Band6 provides the temperature at the top part and along the spicules. As for the \ion{Mg}{2} k line (not shown in this plot), it forms over a broad range of heights so for reasons of clarity we do not show a $\tau=1$ line in this figure. Its formation height has strong wavelength dependence as we show in the next section \S~\ref{results}. \rm We discuss the diagnostic information from \ion{Mg}{2} k with $\tau=1$ geometric heights at different wavelength positions in comparison to the formation height of the other observables for spicules in Paper 2 (morphology model)\rm.

\section{Results}\label{results}


The analysis in the previous sections and the qualitative comparison between observational and synthetic $x-t$ plots reveals that our joint \emph{IRIS} and \emph{ALMA} observations have likely captured a Type-II spicule observed on-disk. Figure~\ref{FIG_RBE} provides strong evidence of the spicule's RBE signatures along its slender structure. Also, the dynamic evolution is characterized by upward and downward motions projected in the plane of the sky (e.g., parabolic profile in Figure~\ref{FIG_3}) as also typically seen even for spicules at the limb. Here, we discuss the similarities in the dynamics of the spicules between the synthetic observables and the observations. 

To address the similarities with respect to the dynamic evolution of the spicules, we produce a set of $x-t$ plots at different wavelength positions for the simulated region of spicules (Figure~\ref{FIG_XT_SIM}) to compare against the $x-t$ plots from the \emph{IRIS} and \emph{ALMA}/Band6 observations which were presented in Figure~\ref{FIG_3}. In addition, we degrade the spatial resolution of the synthetic observables via convolution with the appropriate 2D gaussian kernel to match the observed \emph{ALMA}/Band6 beam size (we pick a ``worst beam size'' value of 0$\farcs$8) and the spatial resolution of the observed \emph{IRIS} rasters (0$\farcs$16 pixel$^{-1}$ along the slit). Also, we produce wavelength-integrated $x-t$ plots as done in Figure~\ref{FIG_3}. The qualitative similarities seen in this comparison are striking. We note, however, that the signatures of spicule 2 are not as clear, perhaps because it is adjacent to spicule 1, therefore it is not seen against a dark background like spicule 1. 

Regarding \emph{ALMA}/Band6, we can see the increase of $T_b$ at the onset of growth of spicule 1 at $(x,t)=($2\,Mm, 3,300\,s$)$ and for spicule 2 around $(x,t)=($1\,Mm, 3,300\,s - 3,400\,s$)$ (display saturated at $T_b$ = 6,000\,K). In the observations (Figure~\ref{FIG_3}), a similar increase in $T_b$ can be seen to develop at the onset of growth at $\sim$16:10 UT. Later, at the moment of the network-jet-like brightening in the simulation at $(x,t)=($2\,Mm, 3,650\,s$)$, we see an enhancement in $T_b$ (same $(x,t)$ coordinates). Even though in the observations the network jet occurred during the calibration gap between \emph{ALMA}/Band6 Scans 13 and 16, there is evidence that a temperature enhancement was still ongoing at $(x,t)=($3$\arcsec$, 16:19 UT$)$. Note, however, that we are comparing the observations to simulated events which are not meant to simulate a specific event, therefore differences might be in the details in the simulation, namely, the amount of current, field configuration, etc. Also, the range of values for the synthesized \emph{ALMA}/Band6 $T_b$ for each of the simulated spicules is about 2,000\,K lower than the observed \emph{ALMA}/Band6 temperatures in the spicules, suggesting that the energy balance in the simulations does not fully capture all the relevant processes (radiation, heating, ionization, etc.)

To identify the geometric heights where \ion{Mg}{2} emission forms in our simulation we use $x-t$ plot of $\tau=1$ heights as a function of wavelength (Figure~\ref{FIG_XT_SIM}; each $x-t$ at a specific wavelength and arranged in velocity space). We also plot the parabolic profile for spicule 1 and 2 (yellow and green dotted lines) as taken from the corresponding $\tau=1$ plots. We typically see that spicules first appear in blueshifts (as high as -37\,km s$^{-1}$), which is consistent with the ascending phase in their evolution. Indications for RBE effects are seen in the initial growth phase of the spicules in both \ion{Si}{4} and \ion{Mg}{2} k (i.e., see blue oval at -25\,km s$^{-1}$ for spicule 1 and at -12\,km s$^{-1}$ for spicule 2; compare areas in blue ovals against the red ovals to see difference in intensities). \ion{Mg}{2} k progressively samples the spicule at maximum elongation at the line rest wavelength (+ 0\,km s$^{-1}$) and in the receding phase of the spicules in the red wing (here, down to +25\,km s$^{-1}$). In the $x-t$ from \emph{IRIS} observations of \ion{Mg}{2} k rest -55\,km s$^{-1}$ (shown in Figure~\ref{FIG_3}), we also get signal at the blue wing $(x,t)=($0-3$\arcsec$, 16:10-16:20 UT$)$. 

However, while the $x-t$ plots of $\tau=1$ in Figure~\ref{FIG_XT_SIM} suggest that we are observing the spicule as it grows, stalls and recedes, the \ion{Mg}{2} k intensity is (1) too weak and (2) the various contributors to the line too complex in the $x-t$ plot to highlight a rough parabolic profile. We emphasize that this is very similar to what is observed. For instance, see areas pointed by arrows where intensity is low (due to absorption, as we show in the following sections). In the  wavelength-integrated map of \ion{Mg}{2} k line, we do not see a full (or even partial) parabolic trace as well-defined as in \emph{ALMA}/Band6 and \ion{Si}{4}. The latter is the case for both the observations (Figure~\ref{FIG_3}) and the simulation. Despite that, during the time of the network jet in the simulation, we see significant intensity enhancement at the \ion{Si}{4} rest wavelength all the way to the blue wing (i.e., area inside dotted circle at -25\,km s$^{-1}$) but also indications of \ion{Mg}{2} absorption in the red wing (arrow in dotted circle at +12\,km s$^{-1}$). Likewise, we see similar enhancement from the \ion{Mg}{2} k line core to the blue wing in the observed $x-t$ (between times 16:17-16:20 UT). 

The most striking resemblance between the simulation and the observations is found in the \ion{Si}{4} $x-t$ (Figure~\ref{FIG_XT_SIM}). High \ion{Si}{4} emissivity has been observed to emanate from around the tip of the spicule as it grows, until it reaches the apex of the $x-t$ parabolic profile where the emission is then seen mostly at the core of the line \citep{DePontieu_etal_2017b,Chintzoglou_etal_2018}. Remarkably, a network jet brightening here also occurs during the descending phase of the $x-t$ profile (dotted circle in selected panels in Figure~\ref{FIG_XT_SIM}) and the emission is seen in redshifts (all the way up to +25\,km s$^{-1}$), as discussed above. Interestingly, the network jet-like brightening appears in \ion{Mg}{2} k in blueshifts and also at the rest wavelength of the line but as a dark feature in the +12 and +25\,km s$^{-1}$ line positions (compare the dotted circle and locations pointed by arrows). \bf We explore this clear intensity depression (in \ion{Mg}{2}) in the spicules as well as anti-correlations with \ion{Si}{4} and \emph{ALMA}/Band6 in Paper II. \rm

To better illustrate the similarities of the spicule in the model and in the observations, we show in Figure~\ref{FIG_TRICOLOR} a tri-color combination of the $x-t$ plots that summarize the evolution seen in observations and in the simulation in \emph{ALMA}/Band6 (red), \ion{Mg}{2} k, (blue; 0.7\,\AA\ integration), and \ion{Si}{4} (green; 0.2\,\AA\ integration). At the same time this also allows us to assess the instantaneous spatial distribution of the heating along the spicular column. Areas where red, green, or blue, colors produce ``color blends'' essentially illustrate the multi-thermality of the plasma. In Figure~\ref{FIG_TRICOLOR}a we show the tri-color plot with the \emph{ALMA}/Band6 data gap masking all observables. In panel b we show all available data together with annotations. The tri-color plot from the observations (Figure~\ref{FIG_TRICOLOR}ab) agrees with the general evolution of the simulated spicule 1 and 2 (Figure~\ref{FIG_TRICOLOR}c). Initially (dashed oval A in Figure~\ref{FIG_TRICOLOR}b), the observed spicular plasma only extends to low heights and appears in relatively lower temperatures (no signal in \ion{Si}{4} emission; however, there is signal in both \emph{ALMA}/Band6 and \ion{Mg}{2} k at the same location producing a magenta color blend). Eventually, the observed spicular plasma achieves higher temperatures (as also manifested by emission seen in \ion{Si}{4} raster scans; e.g., Figure~\ref{FIG_3}), presumably due to the heating from a shock traveling through the growing spicular column (thin ``linear'' bright trace in the $x-t$ plots; Figure~\ref{FIG_TRICOLOR}ab), that leads to both chromospheric and transition region emission. Later on in the observations, shortly after the time when the spicule had reached its maximum elongation, the spicule seems to produce emission in all those wavelength ranges simultaneously. This is marked in Figure~\ref{FIG_TRICOLOR}b with a white arrow as the onset of a network jet brightening. It is worth repeating here that while there is a data gap in \emph{ALMA}/Band6 (dark gap in panel a), there are indications of high signal in \emph{ALMA}/Band6 just after the end of the data gap (bright yellow color blend within dashed oval B). Eventually, as the spicule recedes, significant intensity is found in all three wavelength bands simultaneously (white color blend around 16:22 UT).

The behavior we saw in the observations is also seen in the synthetic tri-color plot for spicules 1 and 2 of Figure~\ref{FIG_TRICOLOR}c; i.e., there is significant intensity at the onset of spicule growth (white dashed ovals A1 and A2 containing the regions with magenta colors), in addition to high emission in \ion{Si}{4}, which here is also seen to ``envelope'' the parabolic profile of the spicule at all times in the $x-t$ plots. Within the time of the network jet brightening (occurring primarily in spicule 2 but also in spicule 1) high intensity is reached in all these wavelengths (area within dashed ovals B1, B2). Therefore, our synthetic observables (Figure~\ref{FIG_4}) compared to the observations in Figure~\ref{FIG_3} support the multi-thermal nature of spicules (e.g., \citealt{Chintzoglou_etal_2018}, as revealed by \emph{VAULT2.0} and \emph{IRIS} observations). However, as we mentioned above, the brightness temperatures for the spicule in the synthetic \emph{ALMA}/Band6 observable are $\approx$2,000\,K lower than $T_b$ in the observations for either of the spicules. 

The consideration of time-dependent ionization in the simulation has the effect of increasing the electron density \citep{Carlsson_Stein_2002, Wedemeyer_etal_2007} in chromospheric heights in the domain as compared to previous simulations that did not consider non-equilibrium ionization effects (e.g., see \citealt{Martinez-Sykora_etal_2020a}). The above also suggests that, with higher electron densities in the chromospheric plasma, \emph{ALMA}/Band6 (sensitive to free-free emission from chromospheric electrons) is probing spicules in optically thick emission that originates from greater heights due to the increased electron density. This can be seen in the $\tau=1$ height shown in Figure~\ref{FIG_5}d (red line), which roughly delineates the contour of the spicules; alternatively, compare the parabolic profile in the $x-t$ plot of \emph{ALMA}/Band6 and \ion{Si}{4} (Figure~\ref{FIG_4}c and e). As we already know from previous studies of spicules seen on-disk (e.g., in \citealt{Chintzoglou_etal_2018}) \ion{Si}{4} emission from Type-II spicules roughly demarkates the latter's linear extents as projected on the plane of the sky.

 In addition, by including all these effects in the simulation, the simulated spicule 2 (and to some extent spicule 1) suddenly brightens along its full length, producing a network jet. The apparent (i.e., simulated plane-of-the-sky) speed of this network jet is $\approx$140\,km s$^{-1}$, a speed more than 3 times the highest Doppler velocities contained in the synthetic spectra of the spicule at that time (and about 6-7 times higher if compared to bulk plasma speeds in the spicule, which at the time of the network jet are primarily downflows). This mismatch between apparent speeds and actual plasma speeds (revealed by RBE/Doppler velocities) supports the idea that some network jets can be rapidly propagating heating fronts \citep{DePontieu_etal_2017b, Chintzoglou_etal_2018}, in contrast with other interpretations, i.e., rapid upward mass motions \citep{Tian_etal_2014}. The similarity between the model and the observations is quite striking and once again points to the impulsive heating and the multithermal nature of spicules \citep{Chintzoglou_etal_2018}.

\section{Summary \& Conclusions}\label{conclusion}

In this work we focused on addressing the nature and the dynamics of chromospheric/transition region jet-like features, i.e., Type-II spicules, using high time-cadence and high spatial resolution data from the \emph{ALMA} and \emph{IRIS} observatories. Our target was a plage region in the leading part of NOAA AR 12651. The spatial distribution of the plage region in the FOV allowed the simultaneous observation of regions of high chromospheric emission in addition to regions with low background emission; the latter permitted the unambiguous on-disk observation of a spicule with \emph{ALMA}/Band6 observations. Thanks to this favorable observing geometry, our analysis was free from many difficulties faced in previous works, such as low signal-to-noise ratio and strong LOS superposition in the data, which is common in observations of spicules at the limb (e.g., \citealp{Tei_etal_2020}). In addition, to assist the interpretation of these unique observations, we employed a 2.5D numerical simulation (Bifrost model) of spicules that considers ambipolar diffusion in non-equilibrium ionization conditions. We produced synthetic observables to compare the model with our observations from \emph{ALMA}/Band6 and \emph{IRIS}.

\begin{figure*}
	\includegraphics[width=\linewidth]{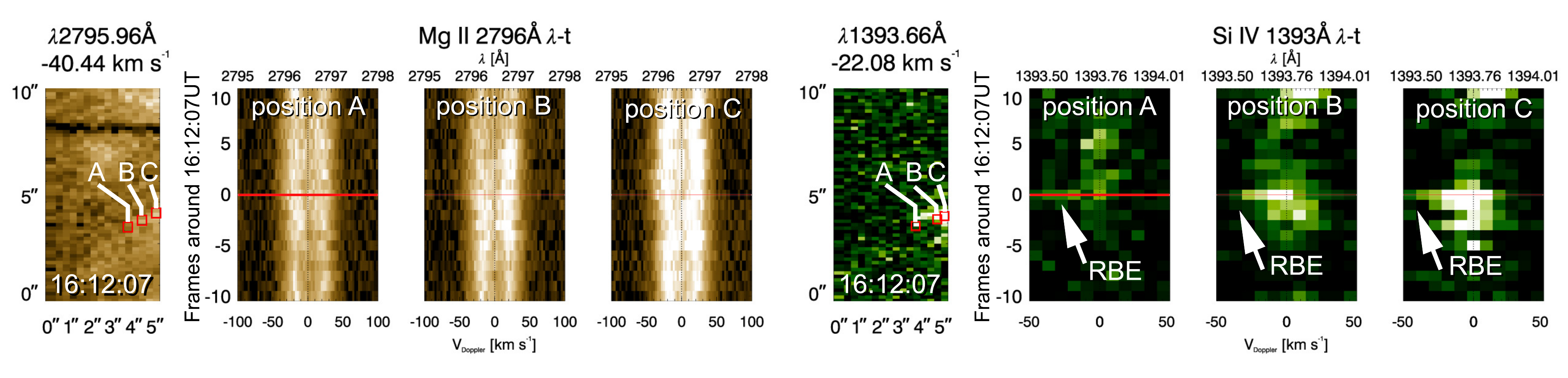}
	\caption{$\lambda-t$ plots for \ion{Mg} {2} and \ion{Si}{4} at three different positions along the linear structure (shown in Figure~\ref{FIG_2} at 16:12 UT), illustrating the RBE nature of the structure with speeds of -50\,km s$^{-1}$ (arrow in Mg II $\lambda-t$; speeds in excess of -50\,km s$^{-1}$ for \ion{Si}{4}). See text for discussion.
	}\label{FIG_RBE}
\end{figure*}

\begin{figure*}
	\includegraphics[width=4.4in]{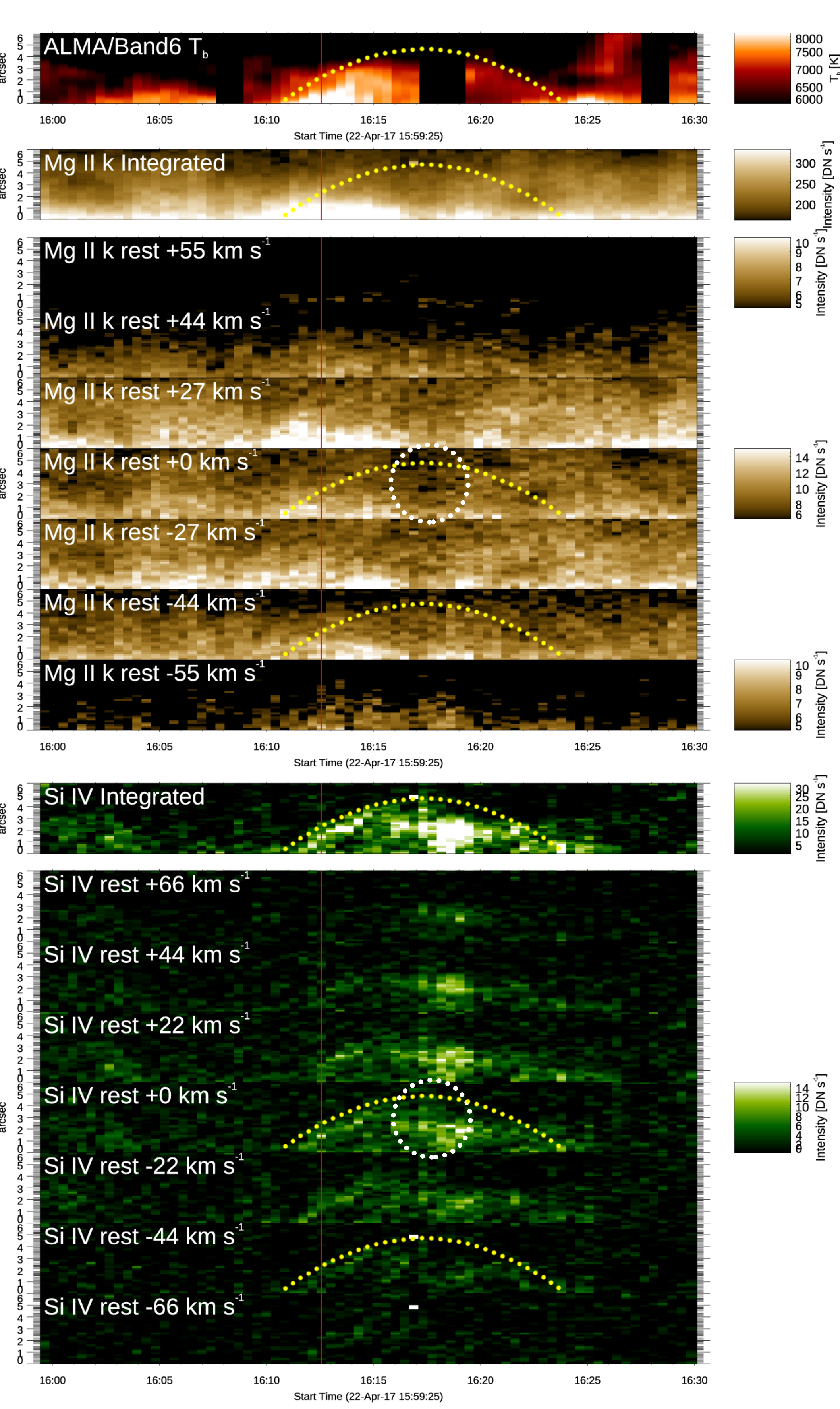}
	\caption{\emph{ALMA}/Band6 and \emph{IRIS} raster $x-t$ plots for the linear-like structure shown in Figures~\ref{FIG_2} and ~\ref{FIG_RBE}. For \emph{IRIS} the stack of $x-t$ plots is arranged in the velocity space as annotated, centered about the rest wavelengths of \ion{Mg} {2} k (middle) and \ion{Si}{4} (bottom), in addition to $x-t$ for wavelength-integrated rasters. The west side of the ``cut'' in Figure~\ref{FIG_2} is at the bottom of the $x-t$ plots. Note the clear parabolic trace in \emph{ALMA}/Band6 and the wavelength-integrated \ion{Si}{4} $x-t$. Also note the enhanced absorption (dotted circle) in \ion{Mg}{2} which alters the appearance of a full parabolic profile. Due to low photon counts in the $x-t$ for \ion{Mg}{2} at $\pm$55 km s$^{-1}$ we reduce the dynamic range as shown in the respective color bars. The red vertical line denotes the time of observations shown in Figure~\ref{FIG_RBE}. See text for discussion.
	}\label{FIG_3}
\end{figure*}

\begin{figure*}
        \includegraphics[width=5.5in]{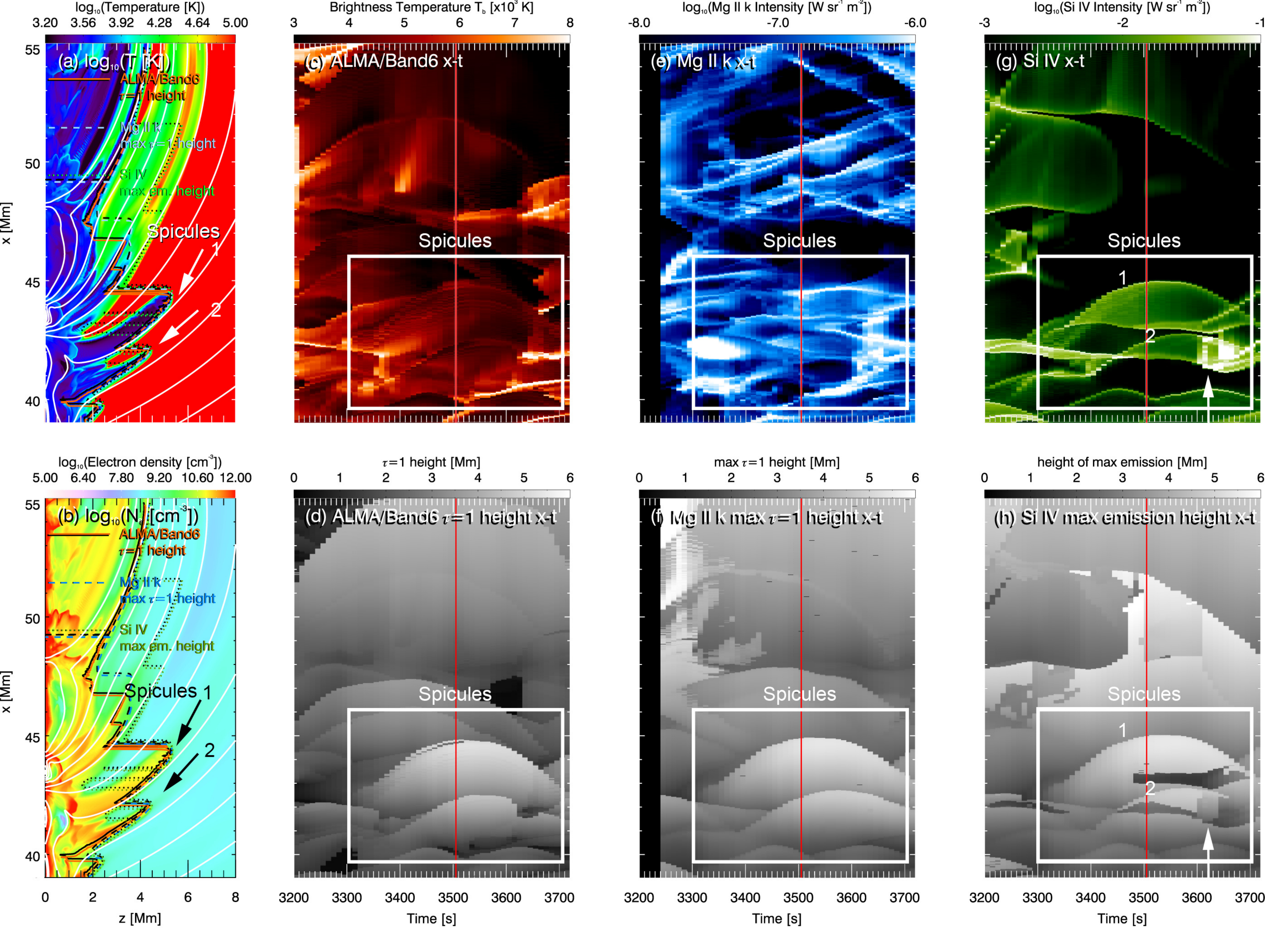}
	\caption{Synthetic $x-t$ plots from the Bifrost simulation showing spicules 1 and 2. Left column panels show a snapshot from the simulation at $t$=3,500\,s for (a) $log_{10}T$ and (b) $log_{10} N_e$. Upper panels (c,e,g) show $x-t$ in \emph{ALMA}/Band6 T$_b$ (in red), \ion{Mg}{2} k (in blue), and \ion{Si}{4} (in green) synthesized intensities, respectively. Bottom panels (d, f, h) present $x-t$ plots for the geometric height where $\tau=1$ for \emph{ALMA}/Band6, maximum $\tau=1$ height for the wavelength-integrated \ion{Mg}{2} k, and the height of maximum emission for \ion{Si}{4}. Boxed areas denote the regions of spicules (best seen in \ion{Si}{4}) and plage. With a red line in the $x-t$ plots we mark the time shown in panels (a) and (b).	
	}\label{FIG_4}
\end{figure*}

\begin{figure*}
	\includegraphics[width=\linewidth]{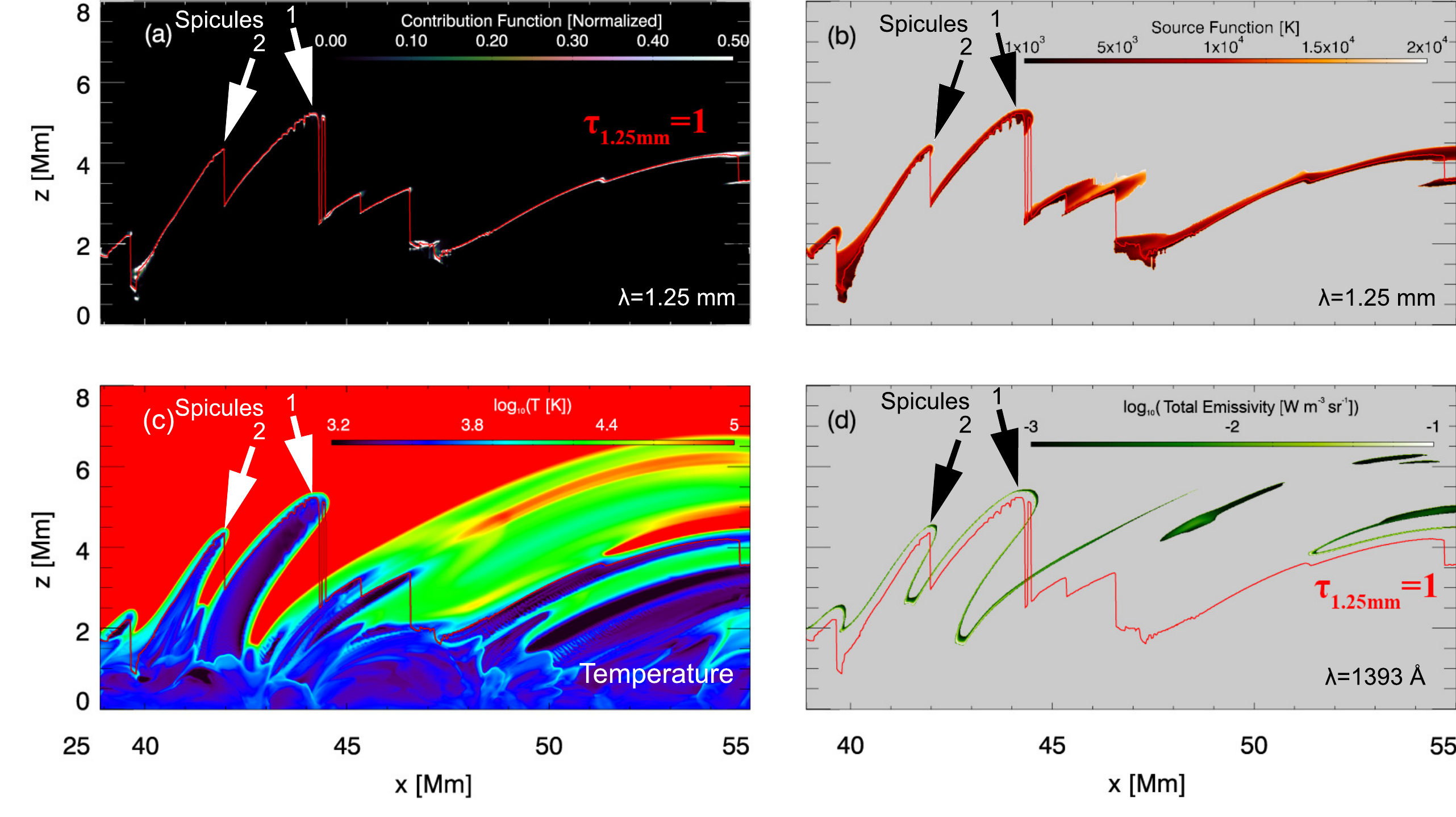}
        \caption{\emph{ALMA}/Band6 1.25\,mm synthesis from the time of the snapshot of the simulation in Figure~\ref{FIG_4} of (a) the normalized contribution function (i.e., $g_\nu/max(g_\nu)$, clipped between 0.0 to 0.5, linear scale) and (b) the related source function, $S_\nu$ (here expressed in units of temperature [K] and shown only for regions where $g_\nu/max(g_\nu)\geq10^{-4}$; gray mask). The red line in all panels shows the \emph{ALMA}/Band6 $\tau=1$ height. (c) Temperature map. (d) Color contour (in green; logarithmic scaling; gray mask clipping low values) of the total emissivity, $\eta$, for \ion{Si}{4} 1393\,\AA. Note the close correspondence in the geometric heights for the total emissivity of \ion{Si}{4} 1393\,\AA\ and $\tau=1$ for \emph{ALMA}/Band6.
} \label{FIG_5}
\end{figure*}

\begin{figure*}
	\includegraphics[width=4.5in]{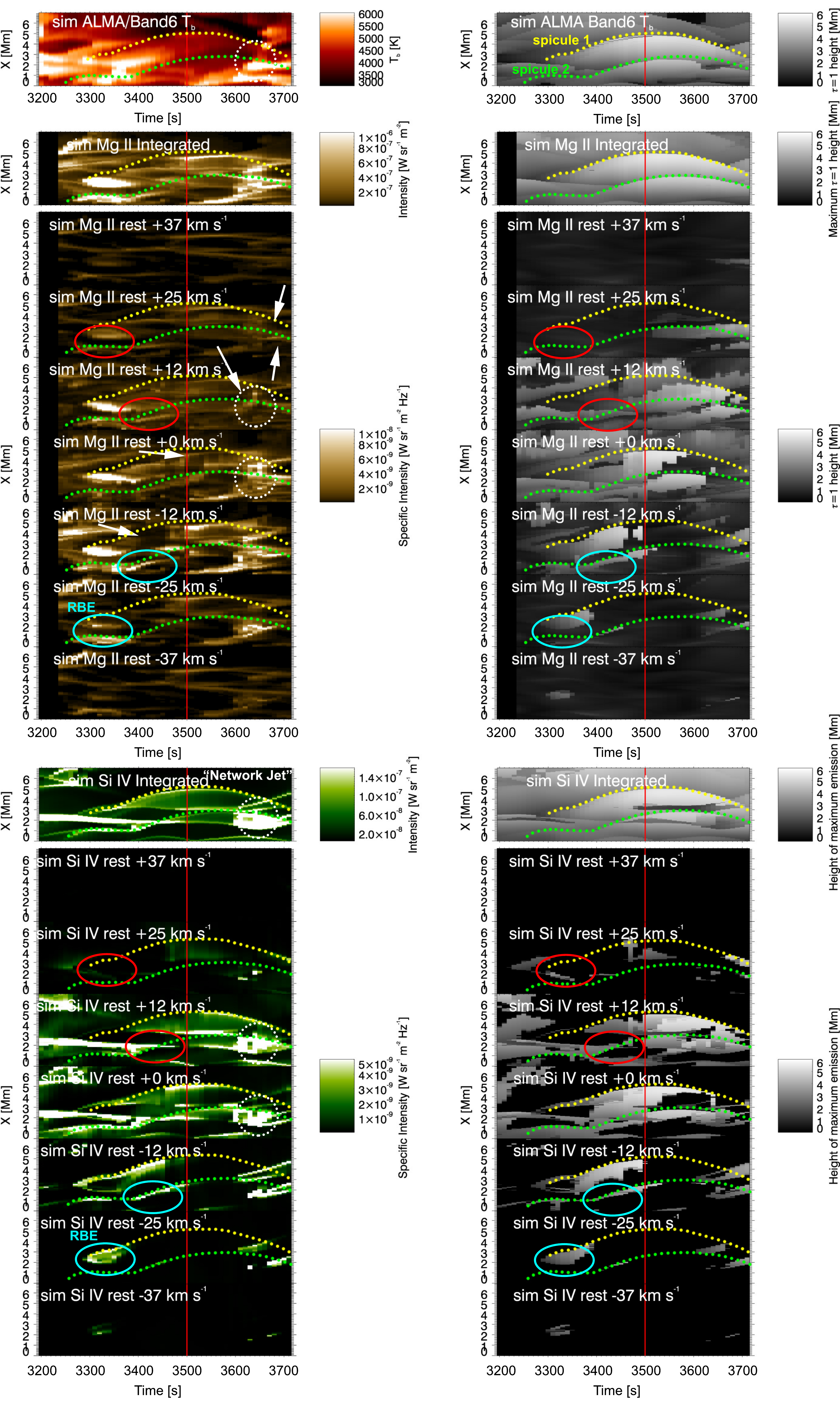}
	\caption{Left panels: $x-t$ plots stacked along wavelength (velocity space) for the synthetic observables produced from the simulation. Note the qualitative similarities with the $x-t$ plots produced from the observations (Figure~\ref{FIG_3}). Right panels: $x-t$ plots of the height where $\tau=1$ for the optically thick observables and the height of maximum emission for the optically-thin \ion{Si}{4}. Dotted lines highlight the parabolic profile of the Type-II spicules 1 and 2 (yellow and green colors, respectively).
	The arrows show locations of enhanced absorption in \ion{Mg}{2} k. The ovals at $\pm$25 and $\pm$12 km s$^{-1}$ pinpoint locations with RBE signatures at the onset of spicule growth. The red vertical line marks the time shown in the previous plots.} \label{FIG_XT_SIM}
\end{figure*}

\begin{figure*}
	\includegraphics[width=4.25in]{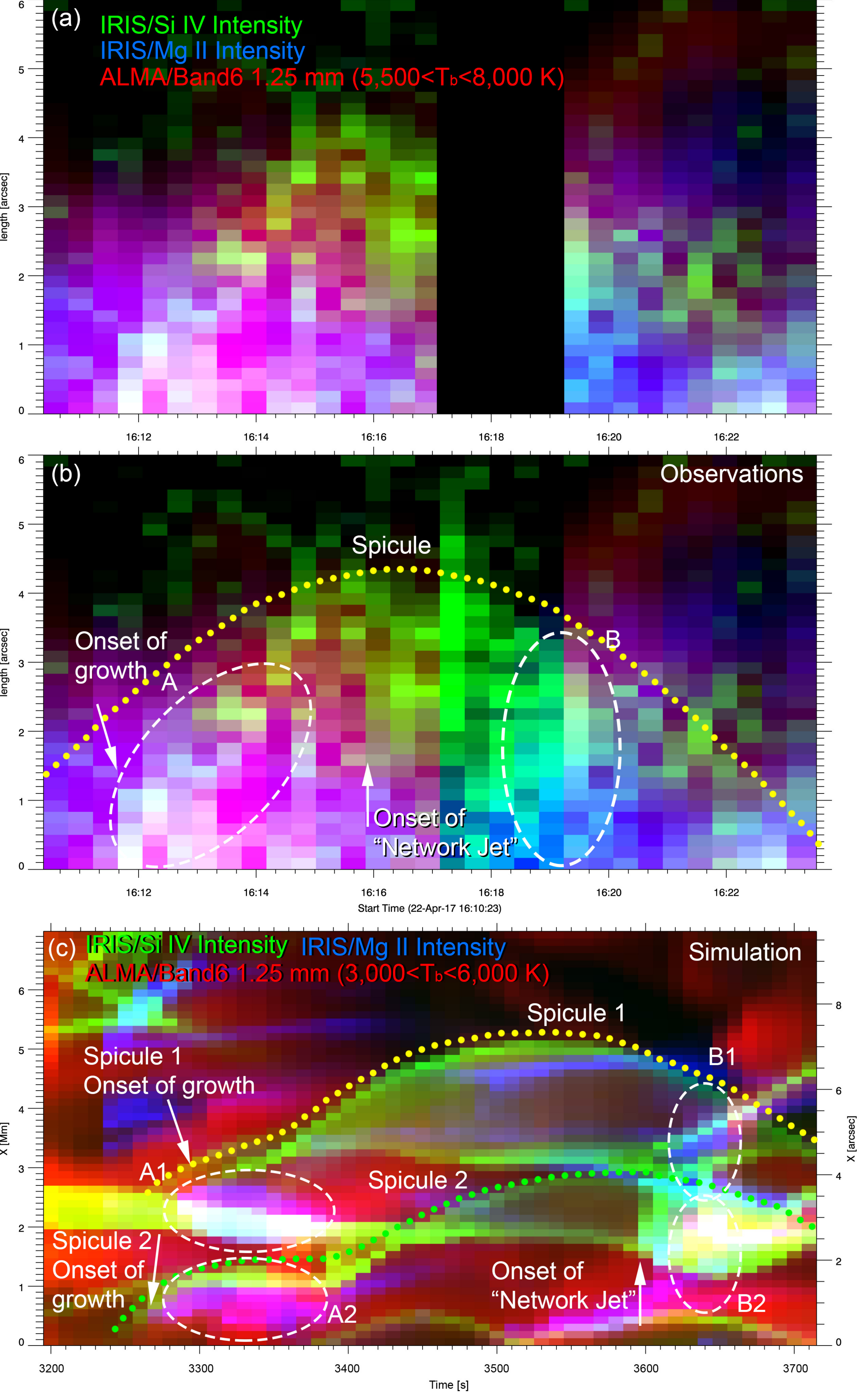}
	\caption{Panels (a) and (b): $x-t$ plot combining observations shown in Figure~\ref{FIG_3} in a tri-color blend: \emph{IRIS} \ion{Mg}{2} (blue), \ion{Si}{4} (green) and \emph{ALMA}/Band6 $T_b$ (red). The \emph{IRIS} \ion{Mg}{2} and \ion{Si}{4} are integrated along $\lambda$ covering the line profiles. Panel (a) shows the \emph{ALMA}/Band6 data gap by blocking \emph{IRIS} data. Panel (b) shows the \emph{IRIS} data at all times. Initially the spicule is seen in low temperatures (magenta color blend). The white arrow marks the time of the sudden network jet brightening of the spicule in \ion{Si}{4} (suggesting $T\approx$80,000\,K). Note, however, that at the same time enhanced emission also comes from spicular plasma in lower temperatures (e.g., \emph{ALMA} $T_b\approx$8,000\,K), suggesting that the spicule is a multithermal plasma structure. Panel (c): similar plot from the corresponding synthetic observables for the simulated spicules of Figure~\ref{FIG_4}. See text for discussion.} \label{FIG_TRICOLOR}
\end{figure*}

Our main findings can be summarized as follows:

\begin{enumerate}
	
\item We conclude that the dynamic linear structure captured in the common \emph{IRIS} and \emph{ALMA}/Band6 FOV is a Type-II spicule. This is supported by (a) the slender profile of the structure, (b) the Rapid Blueshifted Excursion (RBE) in the spectrum of \ion{Mg}{2} k and \ion{Si}{4} from the bulk of the structure (Figure~\ref{FIG_RBE}), (c) the dynamics seen in the space-time plot (parabolic space-time profile; Figure~\ref{FIG_3}) in the observations and in the space-time plots from synthetic observables produced from the simulation (Figure~\ref{FIG_4}).

\item The identified spicule experienced a network jet brightening (apparent speed $\approx$95\,km s$^{-1}$). Our synthetic observables (Figure~\ref{FIG_XT_SIM}) compared with the observations (Figures~\ref{FIG_3} and~\ref{FIG_TRICOLOR}) show a clear agreement with each other, including the occurrence of a network jet brightening in the simulation (apparent speed $\approx$140\,km s$^{-1}$). Note that the \emph{IRIS} raster time-cadence was at 26\,s (while the simulation's time-cadence was at 10\,s), each posing a lower limit in the deduced apparent speed. This apparent brightening in \ion{Si}{4} likely occurred due to a rapidly-propagating heating front along the spicular mass, instead of representing rapid upward mass flows (consistent with \citealt{DePontieu_etal_2017b, Chintzoglou_etal_2018}). In fact, both the observed and the simulated spectra show a clear red-blue asymmetry suggesting mass motions directed away from the observer during the time of the network jet.

\item We confirm the multi-thermal nature of dynamic Type-II spicules \citep{Chintzoglou_etal_2018}. The tri-color space-time plot (combining \emph{ALMA} and \emph{IRIS} observables; Figure~\ref{FIG_TRICOLOR}) reveals a picture strongly suggesting that the spicular plasma emits in multiple temperatures simultaneously, also supported by synthetic observables from the Bifrost simulation (Figure~\ref{FIG_4}).

\item We also noted an interesting intensity depressions highlighted in the  in \ion{Mg}{2} k space-time plots (i.e.,  in the observations in Figure~\ref{FIG_3} and in the simulation in Figure~\ref{FIG_XT_SIM}). We conclude that the apparent anti-correlation or lack of correlation (which may in part be due to observing different parts of the same events/structures either as cospatial or even as non-cospatial owing to LOS projection effects) has its origin in \ion{Mg}{2} opacity effects in plage structures. Strong absorption is the reason behind the low \ion{Mg}{2} intensities emerging from greater geometric heights in the locations of spicules.


\end{enumerate}

In a separate publication (Chintzoglou et al 2020b) we focus on the morphological similarities in plage structures seen between \emph{ALMA} and \emph{IRIS} observations, with an emphasis on the formation height of the optically-thick free-free emission observed with \emph{ALMA}/Band6.

\acknowledgements

This paper makes use of the following \emph{ALMA} data: ADS/JAO.ALMA\#2016.1.00050.S. \emph{ALMA} is a partnership of ESO (representing its member states), NSF (USA) and NINS (Japan), together with NRC (Canada), MOST and ASIAA (Taiwan), and KASI (Republic of Korea), in cooperation with the Republic of Chile. The Joint \emph{ALMA} Observatory is operated by ESO, AUI/NRAO and NAOJ. We gratefully acknowledge support by NASA contract NNG09FA40C (IRIS). JMS is also supported by NASA grants NNX17AD33G, 80NSSC18K1285 and NSF grant AST1714955. VH is supported by NASA grant 80NSSC20K1272.
JdlCR is supported by grants from the Swedish Research Council (2015-03994), the Swedish National Space Board (128/15) and the Swedish Civil Contingencies Agency (MSB). This project has received funding from the European Research Council (ERC) under the European Union's Horizon 2020 research and innovation programme (SUNMAG, grant agreement 759548). 
MS, SJ and SW are supported by the SolarALMA project, which has received funding from the European Research Council (ERC) under the European Union’s Horizon 2020 research and innovation programme (grant agreement No. 682462), and by the Research Council of Norway through its Centres of Excellence scheme, project number 262622.
The simulations and \ion{Mg}{2} synthesis were ran on clusters from the Notur project, and the Pleiades cluster through the computing project s1061, s1630, and s2053 from the High End Computing (HEC) division of NASA. This research is also supported by the Research Council of Norway through 
its Centres of Excellence scheme, project number 262622, and through 
grants of computing time from the Programme for Supercomputing. 
\emph{IRIS} is a NASA small explorer mission developed and operated by LMSAL with mission operations executed at NASA Ames Research center and major contributions to downlink communications funded by ESA and the Norwegian Space Centre. HMI and AIA are instruments on board \emph{SDO}, a mission for NASA’s Living with a Star program.  



\clearpage

\clearpage

\clearpage

\clearpage

\clearpage

\clearpage

\clearpage

\clearpage

\clearpage

\clearpage

\end{document}